\documentclass[12pt]{article}
\usepackage{a4wide}
\usepackage{amssymb}
\usepackage{amsmath}
\usepackage{graphicx}
\usepackage{mathdots}
\usepackage{slashed}
\usepackage{verbatim}
\usepackage{xcolor}

\usepackage{array}
\usepackage{amsthm}

 \numberwithin{equation}{section}

\begin{document}

\begin{flushright}\footnotesize

\texttt{ICCUB-21-002}
\vspace{0.6cm}
\end{flushright}

\mbox{}
\vspace{0truecm}
\linespread{1.1}


\centerline{\LARGE \bf Correlation functions in finite temperature CFT}
\medskip

\centerline{\LARGE \bf  and black hole singularities}

\medskip


\vspace{.4cm}

 \centerline{\LARGE \bf }

\vspace{1.5truecm}

\centerline{
  
    { \bf D. Rodriguez-Gomez${}^{a,b}$} \footnote{d.rodriguez.gomez@uniovi.es}
   {\bf and}
    { \bf J. G. Russo ${}^{c,d}$} \footnote{jorge.russo@icrea.cat}}

\vspace{1cm}
\centerline{{\it ${}^a$ Department of Physics, Universidad de Oviedo}} \centerline{{\it C/ Federico Garc\'ia Lorca  18, 33007  Oviedo, Spain}}
\medskip
\centerline{{\it ${}^b$  Instituto Universitario de Ciencias y Tecnolog\'ias Espaciales de Asturias (ICTEA)}}\centerline{{\it C/~de la Independencia 13, 33004 Oviedo, Spain.}}
\medskip
\centerline{{\it ${}^c$ Instituci\'o Catalana de Recerca i Estudis Avan\c{c}ats (ICREA)}} \centerline{{\it Pg.~Lluis Companys, 23, 08010 Barcelona, Spain}}
\medskip
\centerline{{\it ${}^d$ Departament de F\' \i sica Cu\' antica i Astrof\'\i sica and Institut de Ci\`encies del Cosmos}} \centerline{{\it Universitat de Barcelona, Mart\'i Franqu\`es, 1, 08028
Barcelona, Spain }}
\vspace{1cm}

\centerline{\bf ABSTRACT}
\medskip

We compute thermal 2-point correlation functions in the black brane $AdS_5$ background dual to 4d CFT's at finite temperature for operators of large scaling dimension. We find a formula that matches the expected structure of the OPE. It exhibits an exponentiation property, whose origin we explain. We also compute the first correction to the two-point function due to graviton emission, which encodes the proper time from the event horizon to the black hole singularity.

\noindent

\newpage

\tableofcontents

\section{Introduction}

The study of conformal field theories (CFT's) at finite temperature is of general interest, not only for practical applications --as in real life  critical points are at finite temperature-- but also because finite temperature may reveal new aspects of the theories.
On general grounds, we may explore the theory on different manifolds, 
and a common paradigm is $S^1_{\beta}\times \mathbb{R}^{d-1}$. When the circle of length $\beta$ is assumed to be the euclidean time direction, this is equivalent, provided the appropriate boundary conditions are set, to the CFT at finite temperature $T=\beta^{-1}$. Then, $n$-point functions  probe new important properties not visible on $\mathbb{R}^d$. 

In the following we will focus on the case of the 2-point function of an operator $O$ of dimension $\Delta$, that is, on $\langle O(0)O(\tau,\vec{x})\rangle$. Introducing $|x|^2=\tau^2+\vec{x}^2$, at least in the regime where $|x|\ll \beta$ --~\textit{i.e.} when the distance between the two insertions is much smaller than the size of the thermal circle~-- it is justified to use the OPE. However, unlike  the $\mathbb{R}^d$ case, the size of the $S^1$ provides a scale and allows operators to take a VEV. Thus, the 2-point function unveils an interesting structure which permits to probe physical properties of the CFT.  General features of the thermal 2-point function have been recently discussed in \cite{ElShowk:2011ag,Iliesiu:2018fao}, where it was argued that the 2-point function must admit an expansion in Gegenbauer polynomials, which play the role of conformal blocks on $S^1\times \mathbb{R}^{d-1}$. The coefficients in this expansion can be related to microscopic details of the theory: the structure constants of the OPE, the coefficients of 2-point functions on $\mathbb{R}^d$ and the VEVs which are non-vanishing  on $S^1_{\beta}\times \mathbb{R}^{d-1}$. This structure has been studied in weakly coupled theories including the $O(N)$ model at large $N$ \cite{Iliesiu:2018fao} and free theories \cite{Karlsson:2021duj}. It was also studied using holography \cite{Alday:2020eua} on thermal $AdS$. However, because the stress tensor of the CFT acquires a VEV at finite temperature, the proper gravity dual should be in terms of a black hole in $AdS$. 
Motivated by this, in section 3 we will compute, holographically, 2-point functions in 4d CFT's dual to the black brane $AdS_5$ background.
This computation is closely related to the calculation of 4-point correlation functions of 2 heavy and 2 light operators, discussed in \cite{Fitzpatrick:2019zqz}.
In order to simplify the problem, we will concentrate on operators of large conformal dimension for which we can use the geodesic approximation. The problem then becomes tractable and one can find a systematic expansion for the correlator in terms of the expected Gegenbauer polynomials. The expansion exposes the corresponding coefficients, which are universal for any CFT at strong coupling with a holographic dual. In particular, that corresponding to the energy-momentum tensor block can be compared with the expectations, reassuringly with perfect agreement. One by-product of our analysis is that the leading terms in the $T|x|\ll 1$ and $\Delta\gg 1$ expansion can be resummed into an exponential of the energy-momentum tensor block, a feature also exhibited in the $d=2$ case (and in the free theory in $d=4$, albeit with a different exponentiated block).  Similar  exponentiation property for 4-point functions was observed in CFT$_2$ \cite{Zamolodchikov:1985ie} (see also \cite{Fitzpatrick:2014vua,Besken:2019jyw}) and in CFT$_4$ \cite{Fitzpatrick:2019efk}. For the present case of 2-point functions, we shall see that one can get a  detailed identification of the exponentiated block.

Another important motivation for studying CFT's at finite temperature is that they provide, through holography, a description of black holes.
Indeed, a major challenge in AdS/CFT is understanding how detailed aspects of bulk physics, such as the physics in the black hole interior, are ciphered in CFT correlation functions. Progress in this direction  may shed light on some of the most tantalizing problems in physics, regarding the nature of spacetime singularities and other fundamental issues of quantum gravity. This problematic was undertaken since the beginning of AdS/CFT and, since then, many important advances have been achieved (see {\it e.g.} \cite{Balasubramanian:1999zv,Louko:2000tp,Kraus:2002iv,Fidkowski:2003nf,Hamilton:2006fh,Balasubramanian:2011ur,Heemskerk:2012mn,Papadodimas:2012aq,Hartman:2013qma,Fitzpatrick:2019efk,Fitzpatrick:2020yjb,Dodelson:2020lal,Grinberg:2020fdj}). 

Recently, Grinberg and Maldacena \cite{Grinberg:2020fdj} suggested that geometrical quantities such as the proper time from the event horizon to the singularity can be read off from the behavior of thermal expectation values of CFT operators. It is natural to wonder whether higher-point functions encode, in a similar manner, information about the singularity. This question has, of course, been asked in the past for 2-point functions \cite{Kraus:2002iv,Fidkowski:2003nf}, where it was shown that indeed 2-point functions contain, in a subtle way, information about the interior of the black hole. 

The observation in \cite{Grinberg:2020fdj} that one-point functions are sensitive to the black hole interior stems from the fact that the bulk theory contains a higher derivative coupling between the bulk field $\phi$ dual to $O$ and $W^2$ --~the square of the Weyl tensor~-- which is non-zero on black brane backgrounds. This term is biggest in the vicinity of  the singularity and the main contribution to the one-point function comes
from this region.
As a result, the one-point function of $O$ is proportional to $e^{-i\Delta\,\tau_s}$, being $\tau_s$ the proper time to the singularity. Thus
one can read geometrical properties of the black hole interior just by looking
at thermal 1-point functions in the CFT.
It is natural to expect that thermal 2-point functions, which have a more complicated structure, may contain additional information on the black hole geometry.
Here we shall consider a  model in the 2-point function mediated by a bulk coupling $|\phi|^2W^2$ for four-dimensional CFT's. We assume a $U(1)$ symmetry, which implies that
the first higher-derivative coupling must be of the suggested form.
An example is a bulk field $\phi$ representing a chiral primary operator which carries $U(1)_R$ charge. When $\phi$ is a bulk field of large mass, the corresponding operator has large conformal dimension and we can use the geodesic approximation. 
Then, as we shall see, a similar mechanism as in Grinberg-Maldacena holds in the present case, by which the correlation function at large distances seems to explore the vicinity of the singularity,

The organization of this paper is as follows. In section \ref{2-point} we review the computation of 2-point functions in the geodesic approximation focusing on the $T=0$ case, in particular re-formulating well-known results in a language which will be more suited for the $T\ne 0$ case. In section \ref{finiteT} we then turn  to correlators at non-zero temperature. After reviewing some relevant facts from \cite{Iliesiu:2018fao}, we proceed to compute the 2-point correlation function at non-zero $T$. We first do so for operators inserted at equal $\vec{x}$, \textit{i.e.} correlators with only (euclidean) time dependence, finding in particular the announced  exponentiation. In order to determine the full space dependence, we study the same correlator with general $\vec{x}$- and $\tau$-dependence in perturbation theory in a small temperature expansion. This allows us to explicitly identify the general expansion in terms of Gegenbauer polynomials. As a by-product, we  cross-check our result with the coefficient of the energy-momentum tensor block, finding perfect agreement. The result also shows that 
the exponentiation is of the full energy-momentum tensor block. We then show that this exponentiation takes place as well in CFT$_2$ (for completeness, we review the computation of the 2-point correlator in CFT$_2$ through the geodesic approximation in appendix \ref{CFT2}). In section \ref{n-point} we include the higher-derivative interaction $\bar\phi\phi W^2$ and argue that, in a suitable regime, the correction due to the interaction permits to read
the proper time to the singularity. It also exhibits  resonances effects in graviton emission due to normal frequency modes. Finally, we conclude in section \ref{conclusions} with a summary and comments.

\section{Correlation functions from geodesics}\label{2-point}


On general grounds, the correlation functions of a CFT with a holographic dual can be systematically constructed in an expansion in Witten diagrams, where the vertices can be directly read off from the lagrangian of the bulk theory. In turn, the propagators --either bulk-to-boundary for the ``external legs" or bulk-to-bulk for ``internal lines"-- are constructed by solving the appropriate equations for the Green's functions. As it is well known, for operators of large conformal dimension, such propagators can be approximated by the exponential of minus the geodesic length between the corresponding points (see for instance \cite{Balasubramanian:1999zv,Louko:2000tp}). Then, the Witten diagram looks like a collection of geodesic arcs meeting at the vertices, which are integrated over all the bulk geometry. This picture can be extended as well to 2-point functions of an operator $O$ with conformal dimension $\Delta$. 
They are described by geodesics which penetrate the bulk departing from a boundary point $x_b^{(1)}$ and ending at another boundary point $x_b^{(2)}$. However, it is possible to regard this single geodesic as the junction of two geodesic arcs, one from $x_b^{(1)}$ to some bulk point $x$ and another from $x$ to $x_b^{(2)}$ upon integration over the junction point $x$, \textit{i.e.}

\begin{equation}
\label{correlator}
\langle O(x_b^{(1)})\,O(x_b^{(2)})\rangle=\int_{\rm bulk}dx\, G(x_b^{(1)},x)\,G(x_b^{(2)},x)\,,
\end{equation}
where $G(x_b^{(i)},x)=e^{-iS_{\rm os}(x_b^{(i)},x)}$, being $S_{\rm os}(x_b^{(i)},x)$ the product of the conformal dimension $\Delta$ times the (lorentzian) length  of the geodesic from $x_b^{(i)}$ to $x$; or, equivalently, the on-shell action for a particle of mass $m$ ($\sim \Delta$) which travels between $x_b^{(i)}$ and $x$.\footnote{
The computation of the correlator requires to extracting a factor of $\epsilon^{2\Delta}$, where
$\epsilon$ is a regulator representing the distance to the boundary, 
so that the $\epsilon\rightarrow 0$ limit is well defined.
} Moreover, in the limit of large conformal dimensions, the integral over the junction point $x$ can be done through a saddle point approximation, which significantly simplifies the computation of the 2-point correlation function. 

\subsection{The zero temperature case}

As a warm-up, let us review the computation of 2-point correlation functions at zero temperature in $AdS_{d+1}$ \cite{Dobashi:2002ar,Dobashi:2004nm,Janik:2010gc} (see also \cite{Klose:2011rm,Buchbinder:2011jr,Minahan:2012fh} for applications to 3-point functions). The (euclidean) metric is

\begin{equation}
    ds^2=\frac{R^2}{z^2}\,\big(d\vec{x}^2_{d}+dz^2\big)\,.
\end{equation}
Let us first construct each of the $G(x_b^{(i)},x)$ entering in \eqref{correlator}. To that matter, using the $SO(d)$ symmetry in the euclidean geometry, we consider a particle of mass $m$ moving along the $x$ direction and penetrating in $AdS_{d+1}$. Note that the usual formula relating the conformal dimension to the mass, in the limit of large dimension, becomes $\Delta=mR$. Then, in the gauge where the trajectory is parametrized by $z$, the action is 

\begin{equation}
S= -i\,\Delta\,\int dz\,z^{-1}\,\sqrt{1+\dot{\vec{x}}^2}\, ,
\end{equation}
where dots stand for $z$-derivative.
Note that $ip_x$, the momentum canonically conjugated to $x$, is conserved. The on-shell action is then obtained by integrating $S$ up to a point $z$ --which will eventually be the point where the two arcs meet. One finds

\begin{equation}
\label{Sosp}
    S_{\rm os}= i\Delta\,\log\Big[\frac{\epsilon}{2\Delta\,z}\,\big(\Delta+p_x\sqrt{\frac{\Delta^2}{p_x^2}-z^2}\big)\Big]\,.
\end{equation}
In turn, the conservation of $p_x$ gives a first order equation that can be easily solved:

\begin{equation}
\label{sol}
x=x_1+\frac{\Delta}{p_x}- \sqrt{\frac{\Delta^2}{p_x^2}-z^2}\,.
\end{equation}
Here we have imposed the boundary condition that at the boundary $z=0$, $x(0)=x_1$. 
Solving this equation for $p_x$, one has

\begin{equation}
\label{px}
p_x=2\Delta\frac{(x-x_1)}{(x-x_1)^2+z^2}\,.
\end{equation}
This allows us to evaluate the action $S$ from the boundary point (regulated at $z=\epsilon$) $x=x_1$, up to a generic $z$. We obtain  (we now restore  the full $\vec x_d$ dependence by using $SO(d)$ symmetry)

\begin{equation}
\label{Sos}
S_{\rm os}= i\Delta\log\Big[\frac{z\,\epsilon}{(\vec{x}-\vec{x}_1)^2+z^2}\Big]\,.
\end{equation}
Then $G(\vec{x}_1,x)\approx e^{-iS_{\rm os}}$. Finally, substituting this into \eqref{correlator}, we get

\begin{equation}
\langle O(-\vec{x}_1)\,O(\vec{x_1})\rangle=\int_{AdS_{d+1}}  \frac{dz}{z^d} \, d^dx\  e^{\Delta\log\Big[\frac{z\,\epsilon}{(\vec{x}-\vec{x}_1)^2+z^2}\Big]+\Delta\log\Big[\frac{z\,\epsilon}{(\vec{x}+\vec{x}_1)^2+z^2}\Big]}\,.
\end{equation}
where $(\vec{x},z)$ is interpreted as the joining point over which we have to integrate. Since $\Delta\gg 1$, we may use the saddle point approximation.\footnote{Note that the argument of the logarithms is always smaller than 1. This ensures that, as $\Delta\gg 1$, the exponent goes to $-\infty$.} It is easy to see that the saddle point lies at $\vec{x}=0$ and $z=|\vec{x}_1|$. Note that $\vec{x}=0$ actually follows from symmetry: having chosen the boundary points at $\pm\vec{x}_1$, the joining point will be at the symmetric point $\vec{x}=0$ at some value of $z$. Using the above solution for the saddle-point equations, it follows that 

\begin{equation}
\label{correlatorT=0}
\langle O(\vec{x})\,O(\vec{y})\rangle=\frac{1}{|\vec{x}-\vec{y}|^{2\Delta}}\,,
\end{equation}
just as expected.

\subsection{An alternative derivation}\label{V2}

In the previous computation, the on-shell action \eqref{Sosp} was obtained by using
the conservation of $p_x$, where $p_x$ is to be understood as the momentum corresponding to a particle travelling along the geodesic arc from a boundary point $x_1$ to a generic bulk point $(x,z)$. 
Both $(x_1, 0)$ and $(x,z)$ are to be understood as boundary conditions for our geodesic (where it starts --which will be taken as fixed-- and where it ends). By inverting  the solution to the equation of motion in \eqref{sol}  we can write $p_x$ in terms of $x_1$ and $(x,z)$ to obtain \eqref{Sos}. In this way, $S_{\rm os}$ is the action for the particle with fixed boundary conditions at one end $x=x_1$ at the boundary.
Then, it is the evaluation of the integration by the saddle-point method what fixes the joining point of the two geodesic arcs.

As we will see, the explicit inversion of the solution to the equation of motion to find the analog of \eqref{px} is  not possible in the black brane background. Nevertheless, one can still determine the joining point $(x,z)$
by saddle-point equations. To see this, let us write the total action  as 

\begin{equation}
    S_{T}=i\Delta\,\log\Big[\frac{\epsilon}{2\Delta\,z}\,\big(\Delta+\sqrt{\Delta^2-(p_x^{(1)})^2\,z^2}\big)\Big]+i\Delta\,\log\Big[\frac{\epsilon}{2\Delta\,z}\,\big(\Delta+\sqrt{\Delta^2-(p_x^{(2)})^2\,z^2}\big)\Big]\,;
    \nonumber
\end{equation}
where $p_x^{(i)}$ refers to the momentum associated with the geodesic from $(-x_1,0 )$ to $(x,z)$ and to the geodesic from $(x_1,0)$ to $(x,z)$ respectively. These are, in principle, to be regarded as functions of the boundary conditions as

\begin{equation}
    p_x^{(1)}=p_x^{(1)}(-x_1;(x,z))\,,\qquad p_x^{(2)}=p_x^{(2)}(x_1;(x,z))\,.
\end{equation}
Due to the symmetric choice of the boundary conditions, the joining point must lie at $x=0$. Thus, we can use \eqref{sol} to write equations for the coordinate $z$ of the joining point,

\begin{equation}
\label{solsz}
0=-x_1+\frac{\Delta}{p^{(1)}_x}-\sqrt{\frac{\Delta^2}{(p_x^{(1)})^2}-z^2}\,,\qquad 0=x_1+\frac{\Delta}{p^{(2)}_x}- \sqrt{\frac{\Delta^2}{(p_x^{(2)})^2}-z^2}\,.
\end{equation}
This shows that  $p_x^{(i)}$ are only functions of $z$. Moreover, it is clear that $p_x^{(2)}=-p_x^{(1)}$ (this could be anticipated from the fact that the total momentum after joining the two arcs must be conserved). Thus, denoting $p_x^{(1)}=p_x$, we have

\begin{equation}
\label{relevanteqs}
    S_{T}= 2i\Delta\,\log\Big[\frac{\epsilon}{2\Delta\,z}\,\big(\Delta+\sqrt{\Delta^2-p_x^2\,z^2}\big)\Big]\,;\qquad 0=-x_1+\frac{\Delta}{p_x}- \sqrt{\frac{\Delta^2}{p_x^2}-z^2}\,.
\end{equation}

The saddle point equations are $\frac{dS_T}{dz}=0$ and $\frac{dS_T}{dx}=0$. As for the second equation, all the $x$-dependence of $S$ could only arise through $p_x$. However, since the latter are, as argued above, $x$-independent, $\frac{dS_T}{dx}=0$ automatically follows (which is just a check of the consistency of 
the choice $x=0$). On the other hand, for the $z$-equation we have

\begin{equation}
\label{sades}
    \frac{\partial S_T}{\partial z}+\frac{\partial S_T}{\partial p_x}\,\frac{dp_x}{dz}=0\,.
\end{equation}
$\frac{dp_x}{dz}$ can be computed by differentiating the second equation in \eqref{relevanteqs}.
Then \eqref{sades} leads to the following  saddle-point equation:

\begin{equation}
\label{condisad}
   \sqrt{\Delta^2-p_x^2\,z^2}=0\qquad \longrightarrow \qquad z=\frac{\Delta}{p_x}\,.
\end{equation}
Therefore the on-shell version of \eqref{relevanteqs} is 

\begin{equation}
    S_T\Big|_{\rm os}= 2i\Delta\,\log\Big[\frac{p_x\epsilon}{2\Delta} \Big]\,,\qquad 0=-x_1+\frac{\Delta}{p_x}\,.
\end{equation}
Thus, $e^{-iS_T}= \epsilon^{2\Delta}/|2x|^{2\Delta} $, which, upon restoring the full $\vec x$ dependence
by using $SO(d)$ symmetry, reproduces the result in \eqref{correlatorT=0}.

Finally, it is worth noting  that  the location of the saddle point for $z$ could have been anticipated. Since we are computing a very simple ``Witten diagram" where there is no insertion, the resulting geodesic after joining the two arcs must be smooth. This means that $dz/dx$ must vanish at the joining point, so that we have a $U$-shaped geodesic. Thus, immediate inspection of the equations of motion leads to the condition \eqref{condisad}, which yields the joining point at $z=\frac{\Delta}{p_x}$.

\section{Finite temperature}\label{finiteT}

\subsection{Generalities on thermal 2-point functions}\label{review}

At zero temperature the form of the 2-point correlation function of an operator $O$ of dimension $\Delta$ in \eqref{correlatorT=0} simply follows from conformal invariance, up to an overall factor that depends on the normalization of the operators. The situation is much more interesting at non-zero temperature. Considering the CFT at non-zero temperature amounts to putting it on (euclidean) $S^1_{\beta}\times \mathbb{R}^{d-1}$. For operators inserted at points separated by a distance much smaller than the length $\beta$ of the thermal circle one can still use the OPE. A new feature of non-zero temperature is that operators can have a non-vanishing one-point function. As a result, the 2-point function picks contributions from all operators with non-vanishing OPE coefficients with $O$. This makes the structure of thermal 2-point functions very interesting. It turns out that the correlator $\langle O(0)\,O(\tau,\vec{x})\rangle$, within the regime of validity of the OPE $\sqrt{\vec{x}^2+\tau^2}\ll \beta$, admits the general expression \cite{Iliesiu:2018fao} (see also \cite{ElShowk:2011ag})

\begin{equation}
\label{generalformcorrelator}
    \langle O(0)\,O(\tau,\vec{x})\rangle=\sum_{\mathcal{O}\in {\rm OPE}[O\times O]}\frac{a_{\mathcal{O}}}{\beta^{\Delta_{\mathcal{O}}}}\,C^{(\frac{d}{2}-1)}_{J_{\mathcal{O}}}(\eta)\,|x|^{\Delta_{\mathcal{O}}-2\Delta} ,\qquad \eta=\frac{\tau}{|x|}\,,\ \  |x|=\sqrt{\vec{x}^2+\tau^2}\,.
\end{equation}
The sum in \eqref{generalformcorrelator} runs over all operators $\mathcal{O}$ with dimension $\Delta_{\mathcal{O}}$ and spin $J_{\mathcal{O}}$ appearing in the $O\times O$ OPE. Here $C^{(\frac{d}{2}-1)}_{J_{\mathcal{O}}}(\eta)$ are Gegenbauer polynomials, and the combinations $C^{(\frac{d}{2}-1)}_{J_{\mathcal{O}}}(\eta)\,|x|^{\Delta_{\mathcal{O}}-2\Delta}$ can be regarded as conformal blocks on the cylinder. Finally, $a_{\mathcal{O}}$ are numerical coefficients, which can be related to microscopic quantities as

\begin{equation}
    a_{\mathcal{O}}=\frac{f_{OO\mathcal{O}}\,b_{\mathcal{O}}}{c_{\mathcal{O}}}\,\frac{J!}{2^J\,(\frac{d}{2}-1)_J}\,;
\end{equation}
where  $f_{OO\mathcal{O}}$ is the corresponding OPE coefficient, $c_{\mathcal{O}}$ is the normalization of the 2-point function for  $\mathcal{O}$ and $b_{\mathcal{O}}$ is the coefficient of the one-point function of $\mathcal{O}$ (for instance, for scalars, $\langle\mathcal{O}\rangle=b_{\mathcal{O}}\,T^{\Delta_{\mathcal{O}}}$). 

Among the $a_{\mathcal{O}}$ coefficients, that corresponding to the energy-momentum tensor is particularly interesting. Note that the energy-momentum tensor ${\cal T}$ is a dimension $d$ operator of spin 2, and it thus corresponds to $C^{(\frac{d}{2}-1)}_{2}(\eta)$ with $\Delta_{\mathcal{O}}=2$. The OPE coefficient $f_{OO\mathcal{O}}$ is determined by a Ward identity,

\begin{equation}
    f_{OO{\cal T}}=-\frac{d}{d-1}\frac{\Delta}{{\rm vol}(S^{d-1})}\,,\qquad {\rm vol}(S^{d-1})=\frac{2\pi^{\frac{d}{2}}}{\Gamma\big(\frac{d}{2}\big)}\,.
\end{equation}
Moreover \cite{Iliesiu:2018fao}

\begin{equation}
b_{{\cal T}}=\frac{d}{d-1}\frac{\langle T^{00}\rangle_{\beta}}{T^d}\,.
\end{equation}
Hence

\begin{equation}
    a_{{\cal T}}=-\frac{2d\Delta}{(d-1)^2\,(d-2)\,{\rm vol}(S^{d-1})}\,\frac{\langle T^{00}\rangle_{\beta}}{c_{\cal T}\,T^d}\,;
\end{equation}
where $c_{\cal T}$ stands for the coefficient of the energy-momentum tensor correlator normalized as in \cite{Osborn:1993cr}. For future reference, let us quote the $d=4$ value

\begin{equation}
\label{aT}
    a_{{\cal T}}=-\frac{2\Delta}{9\pi^2}\,\frac{\langle T^{00}\rangle_{\beta}}{c_{\cal T}\,T^4}\, .
\end{equation}

On general grounds, thermal 2-point correlation functions must satisfy the KMS condition $\langle O(0)\,O(\tau,\vec{x})\rangle\sim \langle O(0)\,O(\tau+ \beta,\vec{x})\rangle$. Note that \eqref{generalformcorrelator} does not manifestly exhibit such periodicity, which
lies outside the regime of validity of the OPE (see \cite{Iliesiu:2018fao} for further developments). 

\subsubsection{Case study: free scalar fields}

It is instructive to consider in detail the case of a massless free scalar field $\phi$.  In $d=4$ the thermal 2-point function in position space is

\begin{equation}
\label{prop}
\langle \phi(0)\phi(\tau,\vec{x})\rangle=\frac{\pi}{2\beta}\,\frac{1}{|\vec{x}|}\,\Big[\coth\big(\frac{\pi}{\beta}(|\vec{x}|-i\tau)\big)+\coth\big(\frac{\pi}{\beta}(|\vec{x}|+i\tau)\big)\Big]\, ,
\end{equation}
where we have chosen the normalization so that at zero temperature we recover \eqref{correlatorT=0} with $\Delta=1$. Note that \eqref{prop} manifestly satisfies the KMS condition of invariance under $\tau\rightarrow \tau+ \beta $.

In the regime of $\sqrt{\tau^2+\vec{x}^2}\ll \beta$, \eqref{prop} admits the expansion

\begin{equation}
\label{freelowT}
\langle \phi(0)\phi(\tau,\vec{x})\rangle=\sum_{n=0} 2\zeta(2n) \frac{x^{2n-2}}{\beta^{2n}}\,C_{2n-2}^{(1)}(\eta)\,;
\end{equation}
which is precisely of the form given in \eqref{generalformcorrelator} (and it reproduces the result in \cite{Iliesiu:2018fao}). In particular, we recognize the contribution of the operators $[\phi\phi]_{n,\ell}=\phi\partial_{\mu_1}\cdots\partial_{\mu_{\ell}}\,(\partial^2)^n\phi$ for even $\ell$, which was anticipated in \cite{Iliesiu:2018fao}. It should be noted that for a free field theory $\partial^2\phi$ is a redundant operator and thus drops from \eqref{freelowT}. 

The contribution of the energy-momentum tensor can be read from the coefficient multiplying the $C_2^{(1)}(\eta)$ Gegenbauer polynomial and it is $2\zeta(4) =\frac{\pi^4}{45}$.
This is the expected value, since for free fields, 
$$
\langle T^{00}\rangle_{\beta}=-\frac{2(d-1)\zeta(d)}{ {\rm vol}(S^{d-1})}\, T^d\ ,\qquad 
c_{\cal T}=\frac{d}{d-1}\,\frac{1}{\Big({\rm vol}(S^{d-1})\Big)^2}\ .
$$

\subsection{Thermal 2-point functions for heavy operators
from holography}

Let us consider a CFT$_d$ at finite temperature with an holographic description in terms of  the black brane in $AdS_{d+1}$,

\begin{equation}
\label{blackbrane}
    ds^2=\frac{R^2}{z^2}\big(-f(z)\,dt^2+\frac{dz^2}{f(z)}+d\vec{x}^2_{d-1}\big)\,,\qquad f=1-\frac{z^d}{z_0^d}\,,\qquad z_0=\frac{d}{4\pi}\beta\,.
\end{equation}
Using the same recipe as in \eqref{correlator}, we can compute holographically correlators for operators of large conformal dimension by approximating $G(x_b,x)$ by geodesic arcs in the black brane background in \eqref{blackbrane}. From now on, we specialize to the case $d=4$ and we will choose units where $z_0=1$.

\subsubsection{Geodesic arcs}

For non-zero $T$, the background \eqref{blackbrane}  exhibits $SO(d-1)$ symmetry. Using this, 
it will be sufficient to consider the motion of a particle in the geometry \eqref{blackbrane} with $t=t(z)$, $x_1\equiv x=x(z)$ and $x_i=0$, $i\neq 1$. The corresponding (lorentzian) action is

\begin{equation}
S=-\Delta\int dz\,z^{-1}\,\sqrt{f\,\dot{t}^2-\frac{1}{f}-\dot{x}^2}\,.
\end{equation}
It is clear that the canonical  momenta conjugate to $t$ and $x$  are conserved. Let us denote them by $p_t=\Delta\,\mu$ and $p_x=\Delta\,\nu$, respectively. This allows to write the first order equations as follows

\begin{equation}
\label{txeom}
\dot{t}=\mp \frac{\mu \,z}{f\,\sqrt{-f +(\mu ^2-f\, \nu^2)\,z^2}}\,,\qquad \dot{x}=\pm \frac{\nu\,z}{\sqrt{-f +(\mu^2-f\,\nu^2)\,z^2}}\, .
\end{equation}
Then, the on-shell action takes the form

\begin{equation}
S=-i \Delta\,\int dz\,\frac{1}{z\,\sqrt{f\,-\mu ^2z^2+f\nu ^2z^2}}\,.
\end{equation}

The equations of motion in \eqref{txeom} can be integrated explicitly in terms of elliptic integrals. We will restrict to the case of $\nu=0$, where  the integrals
simplify. This corresponds to insertions of operators at equal points on the $\mathbb{R}^{3}$. The relevant equations then become

\begin{equation}
S=-i \Delta\,\int dr\,\frac{1}{2r\,\sqrt{1-r^2\,-\mu ^2r}}\, \ ,\qquad \dot{t}=\mp \frac{i}{2} \frac{\mu }{(1-r^2)\,\sqrt{1-r^2 -\mu ^2\,r}}\,
,  \label{segunda}
\end{equation}
where $r=z^2$.
Note that close to the boundary $\dot{t}\rightarrow \mp\,i\, \frac{\mu}{2}$. Thus, for real $\mu$, $t$ is imaginary, recovering the Wick rotation (\textit{c.f.} the $T=0$ case, where we worked directly in the euclidean, and hence the momentum was purely imaginary).

Integrating the action from the boundary at $z=\epsilon$ up to a generic bulk point $z$, we find

\begin{equation}
    S=\frac{1}{2} i\Delta\, \log \left(\frac{\epsilon^2 \left(2- \mu ^2 z^2+2
   \sqrt{1-\mu ^2 z^2-z^4}\right)}{4 z^2}\right)\,.
\label{nuevax}
\end{equation}
In turn, the solution to the equation of motion with boundary conditions $t(0)=t_1$ is

\begin{eqnarray}
    t-t_1 &=&-\frac{1}{4} \log \left(\frac{\mu ^2+\left(\mu ^2+2\right) z^2+2 i \mu 
   \sqrt{-z^4-\mu ^2 z^2+1}-2}{\left(\mu ^2+2 i \mu -2\right)
   \left(1-z^2\right)}\right)
   \nonumber\\
   &-&\frac{1}{4} i \log \left(\frac{\mu ^2-\left(\mu
   ^2-2\right) z^2+2 \mu  \sqrt{-z^4-\mu ^2 z^2+1}+2}{\left(\mu ^2+2 \mu +2\right)
   \left(z^2+1\right)}\right)\,.
\label{tedepex}
\end{eqnarray}
The second geodesic arc is obtained by changing $t_1\to -t_1$ and $\mu\to -\mu $ in \eqref{tedepex}.

\subsubsection{Two-point function with time dependence}
\label{secciontresdos}

We can now compute the 2-point function $\langle O(t_1)O(-t_1)\rangle$ by joining two geodesic arcs: one from $(-t_1,\vec 0,z=0)$ to the bulk point $(t,\vec x,z)$, the other from the bulk point $(t,\vec x,z)$ to the boundary point $(t_1,\vec 0,z=0)$. Then we integrate over the bulk point where the two arcs are joined. Since \eqref{tedepex} cannot be inverted to find the explicit form of $G(x_b,x)$, we will follow the same steps as in the $T=0$ case in section \ref{V2}. Just like in the $T=0$ case, the saddle point  will be at $t=0$, $\vec x=0$ by symmetry.
 Then, \eqref{tedepex} evaluated at the joining point formally provides an equation for $\mu$ as a function of $\pm t_1$ and of the bulk joining point $z$. As in section \eqref{V2}, the bulk joining point is determined
by the saddle point equation for $z$, which is given by

\begin{equation}
    0=\frac{dS}{dz}=  \frac{\partial S}{\partial z}+\frac{\partial S}{\partial \mu}\frac{d \mu}{d z}\ ,
\end{equation}
where $\frac{d \mu}{d z}$ is read off from \eqref{tedepex} evaluated at the joining point. One obtains
the  equation,
\begin{equation}
    0= \sqrt{1- z^4-\mu^2z^2}\ .
\end{equation}
Similarly to the $T=0$ case, this is the expected saddle-point equation, since it ensures a smooth ($U$-shaped) geodesic, {\it i.e.} it corresponds to the turning point where $dz/dt$ vanishes. The solution is
\begin{equation}
\label{saddlez}
    z_{\pm}^2=\frac{-\mu^2\pm\sqrt{4+\mu^4}}{2}\,.
\end{equation}
The minus sign solution is complex, so we keep the plus sign, which gives us $z_{\star}=z_+\in (0,1)$, that is, the full geodesic living outside the horizon. Then, the two-point correlation function
is 

\begin{equation}
    \langle O(t_1) O(-t_1)\rangle = e^{-2iS\big|_{\rm os}} =
   4^{-\Delta} \left(\mu ^4+4\right)^{\frac{\Delta}2}\ ,
\label{correit}
\end{equation}
modulo a multiplicative numerical constant which we will neglect (we may fix it \textit{a posteriori} by the normalization of the $T\rightarrow 0$ limit of the 2-point function).

In order to find the time dependence of the correlator, we should trade $\mu$ in \eqref{correit} by its expression in terms of $t_1$. This can be in principle read off from \eqref{tedepex} evaluated at the joining point at $t=0$ and $z=z_{\star}$. This gives the following expression:
\begin{equation}
  \tau_1 \equiv i t_1= - \frac{1}{4}  \log \left(\frac{(\mu -2) \mu +2}{\sqrt{\mu ^4+4}}\right)+\frac{i}{4}
   \log \left(\frac{\sqrt{\mu ^4+4}}{\mu  (\mu +2 i)-2}\right)\ ,
    \label{tmux}
\end{equation}
where we have introduced the euclidean time $\tau_1$. The formulas \eqref{tmux} and \eqref{correit} implicitly define the two-point function $\langle O(\tau_1) O(-\tau_1)\rangle $. First discussed in \cite{Fidkowski:2003nf} (note that our $\mu$ is $i\,E$ in that reference), they encode a  lot of information, including details of the black hole interior, when explored in the complex $\mu$ plane.

In the following we will be interested on real $\mu$. 
Note that, for any real $\mu$, the argument of the log in the first term in \eqref{tmux}  is always real and positive. 
In turn, the argument of the second log is a pure phase $e^{i \theta}$.
%
%
Because of the different branches of the log, $\theta$ is defined up to multiples of $2\pi$. This implies that a given $\mu$ defines $\tau_1$ up to multiples of $\frac{\pi}{2}\equiv\frac{\beta}{2}$. \footnote{To restore the $\beta$-dependence, note that in $d=4$, $z_0=1=\frac{\beta}{\pi}$.} In particular, it follows that $\langle O(\tau_1) O(-\tau_1)\rangle\sim \langle O(\tau_1-\frac{\beta}{2}) O(-\tau_1+\frac{\beta}{2})\rangle$, which is precisely the KMS periodicity condition.

Let us consider the  small $\tau$ behavior, \textit{i.e.} the limit of nearly coincident points. 
From \eqref{tmux} we find that, for large $\mu$,
\begin{equation}
\label{taumu}
  \tau_1=  \frac{1}{\mu }-\frac{4 }{5 \mu ^5}+\mathcal{O}\left(\frac{1}{\mu^6 }\right)\ .
\end{equation}
This regime thus represents the small $\tau$ limit. Moreover, since we are looking at the large $\mu$ range, this  will  correspond to the dominant solution ({\it c.f.} \eqref{correit}). Computing
higher orders in \eqref{taumu}, we find that, at nearly coincident points, the two-point function exhibits the following behavior 
\begin{eqnarray}
\log\langle O(0) O(\tau)\rangle &=& -2\Delta \log \tau +\Delta \frac{\pi ^4  T^4 \tau ^4}{40} \bigg(
 1+\frac{11}{360} \pi ^4 \tau ^4 T^4+\frac{89 \pi ^8 \tau ^8 T^8}{46800}
 \nonumber\\ 
 &+& \frac{90079
   \pi ^{12} \tau ^{12} T^{12}}{572832000}+\frac{126881 \pi ^{16} \tau ^{16}
   T^{16}}{8353800000}+O\left(\tau ^{20}\right)\bigg)\ ,
\label{logcor}
 \end{eqnarray}
 where we have  restored the factors of temperature. The  terms in the first line agree
 with the terms computed in \cite{Fitzpatrick:2019zqz}. From \eqref{logcor} we obtain

\begin{eqnarray}
 \langle O(0) O(\tau)\rangle &=&   \frac{1}{\tau^{2\Delta}}\bigg(1+\frac{ \Delta\, \pi^4\,T^4\, \tau ^4}{40}+\frac{\Delta\,\pi^8\,T^8\tau ^8}{28800}\,
   (9\,\Delta+22)
   \nonumber\\
   &+&\frac{\Delta\,\pi^{12}\,T^{12}\,\tau ^{12}}{14976000}\, (39\,\Delta^2+286\,\Delta+712)+\mathcal{O}\left(\tau
   ^{16}\right)\bigg)\ .
      \label{geor}
\end{eqnarray}
A particular  large $\Delta $ limit exists, with $T\tau\to 0$ and fixed $\Delta (T\tau)^4$.  This limits keeps only the first subleading term in \eqref{logcor}. Concretely, one  obtains

\begin{eqnarray}
     \langle O(0) O(\tau)\rangle  &=&   \frac{1}{\tau^{2\Delta}}\left(1+\frac{ \pi^4\,(\Delta\,T^4\, \tau ^4)}{40}+\frac{\pi^8\,(\Delta\,T^4\tau ^4)^2}{3200}+\frac{\pi^{12}\,(\Delta\,T^{4}\,\tau ^{4})^3}{384000}+\cdots\right)
     \nonumber\\
     &=& \frac{1}{\tau^{2\Delta}}\, \exp\left[{\frac{ \pi^4\,\Delta\,T^4\, \tau ^4}{40}}\right]\ .
      \label{georresummed}
\end{eqnarray}
It should be noted that in this ``double scaling limit" the correlator does not exhibit the KMS periodicity. This is expected, since the regime of the limit requires small $T\tau$, which clearly is not maintained under $\tau\to\tau+\beta$.

 It is instructive to examine the analog limit for (completely connected) correlators of $O=\phi^n$ in the free theory. The correlator is just the $n$-th power of that in \eqref{prop}. Expanded in powers it reads

\begin{eqnarray}
    \langle O(0)\rangle O(\tau,\vec{x})\rangle &=&
  \frac{1}{|x|^{2n}}\,\Big[1+2\zeta(2)\,n\,T^2\,|x|^2\,C^{(1)}_0(\eta)
   \nonumber\\
  &+&2\,n\,T^4\,|x|^4\,\Big( \zeta(4)C^{(1)}_2(\eta)+(n-1) \left(\zeta(2) C_0^{(1)}(\eta)\right)^2\Big)+\cdots\Big]\,.
\nonumber
\end{eqnarray}
Taking the large $n$ ($\equiv\Delta$) limit of this expression and resumming the result we find

\begin{equation}
    \langle O(0)\rangle O(\tau,\vec{x})\rangle=\frac{1}{|x|^{2n}}\,e^{2\zeta(2)\,n\,T^2\,|x|^2\,C^{(1)}_0(\eta)}\,.
\end{equation}
We note that this limit is dominated by the block associated with $\phi^2$, which corresponds to the $C^{(1)}_0(\eta)$ Gegenbauer polynomial.

\subsubsection{Including $x$ dependence}
\label{qavatar}

Let us now consider the general two-point function   $\langle O(0) O(\tau,\vec{x})\rangle $
for operators of large dimension. It can be computed in the geodesic approximation, just as we did
for $\langle O(0) O(\tau)\rangle $, by integrating \eqref{txeom}. 
However, as the integrals lead to complicated expressions in terms of elliptic integrals, it is more illuminating to consider a small temperature expansion (by conformal invariance, this is equivalent to
a short-distance expansion).
We thus look for geodesics that hit the boundary at $(t_1,\,x_1)$ and  end at the generic joining point which,
by symmetry, is located at $(t=0,\, \vec x=0,\,z)$. In the short-distance regime,
the geodesic penetrates very little into the bulk and remains close to the boundary.
We have already seen in \eqref{taumu} that small $\tau $ corresponds to large $\mu$, where
 $z_{\star}\sim \frac{1}{\mu}$.  Consistently, the result in \eqref{geor} is a series expansion around the $T=0$ case in powers of $T\tau$. 
Since the temperature enters the equations of motion through $z_0^{-4}\sim T^4$, the correlation function will be an expansion in powers of $T^4$. This is of course consistent with \eqref{geor}. 
In general, $\mu$, $\nu$ have temperature dependence and admit an expansion of the form

\begin{equation}
    \mu=\sum_{i=0} \mu_i\,T^{4i}\,,\qquad \nu=\sum_{i=0}\nu_i\,T^{4i}\,.
\label{munus}
\end{equation}

Next, we substitute the expansions \eqref{munus} into the exact solution, 
expand the resulting expression in powers of $T$, and   solve, order by order, for the $\mu_i$ and $\nu_i$ as functions of the specified boundary conditions.  
In this way we construct $G(x_b,x)$ to any desired order in $T$. 
We then substitute these expressions  into the on-shell action. The resulting total on-shell action  still has to be extremized for $z$.  Writing $z=\sum_{i=0} z_i\,T^{4i}$, we can solve the corresponding equation 
for the $z_i$ to the appropriate order in $T$. Finally, the on-shell action evaluated on this $z$
gives  the two-point correlation function. We omit the intermediate expressions which are very lengthy. The final result is given by

\begin{eqnarray}
\label{georxt}
     && \langle O(0) O(\tau,\vec{x})\rangle =
    \\ \nonumber 
     &&  \frac{1}{|x|^{2\Delta}}\Big[1+\frac{ \Delta\, \pi^4\,T^4}{120}\,C^{(1)}_2(\eta)\,|x|^4\,
     +\frac{\Delta^2\,\pi^8\,T^8}{28800}\,
   \Big(C_4^{(1)}(\eta)+C_2^{(1)}(\eta)+C_0^{(1)}(\eta)\Big)\,|x|^8+\cdots \Big]\ .
\end{eqnarray}
where, as earlier, $\eta\equiv \tau/|x|,\ |x|=\sqrt{\vec x^2+\tau^2}$.
As a cross-check, upon setting $\vec{x}=0$ in \eqref{georxt}, we exactly recover \eqref{geor}. 

It is of interest to identify the operators contributing to \eqref{georxt}. The first term clearly corresponds to the energy-momentum tensor, which at non-zero $T$ takes a VEV. The remaining terms correspond to dimension $4n$ and spin $2n, 2n-2,\cdots,0$ operators, which are naturally identified with the different contractions in the powers of the energy-momentum tensor. Furthermore, as a consequence, we can identify $a_{\mathcal{T}}$ in \eqref{aT}. This leads to
the formula:

\begin{equation}
\label{formula}
 -\frac{\langle T^{00}\rangle_{\beta}}{T^4}=\frac{3\pi^6}{80}\,c_{\mathcal{T}}\,.
\end{equation}
%
This is a universal relation which should be valid for any four-dimensional CFT with a gravity dual.
%
We can check it explicitly in the case of $\mathcal{N}=4$ super Yang-Mills theory. Using the formulas in \cite{Osborn:1993cr} for the $\mathcal{N}=4$ field-theory content (in that language, $N^2$ gauge fields, $6N^2$ real scalars and ``$\frac{4}{2}\,N^2$ Dirac fermions"), one finds $c_{\mathcal{T}}=\frac{10\,N^2}{\pi^4}$. Thus \eqref{formula} gives $\langle T^{00}\rangle_{\beta}=-\frac{3\,\pi^2}{8}N^2T^4$, in nice agreement with the known result \cite{Gubser:1998nz}.\footnote{Note that $F=E+T\,dF/dT$, so that if $E=-\langle T^{00}\rangle_{\beta}=-(d-1)\,f\,T^d$, then $F=f\,T^d$.}

Another evidence of the consistency of the formula \eqref{formula} arises in the context of
 SUSY CFT's corresponding to $D3$ branes probing a $CY_3$ cone over a base $H$, when the ten-dimensional near-brane geometry is $AdS_5\times H$. In this example, using the fact that for CFT's with a gravity dual the $a=\frac{\pi^3}{4\,{\rm Vol}(H)}$ central charge equals $c_{\mathcal{T}}$, we can write

\begin{equation}
\label{ratio}
 \frac{\langle T^{00}\rangle_{\beta}}{\langle T^{00}\rangle_{\beta}^{\mathcal{N}=4}}=\frac{{\rm Vol}(S^5)}{{\rm Vol}(H)}\,.
\end{equation}
Let us compare this prediction with the  holographic computation of $\langle T^{00}\rangle_{\beta}$.  The only dependence on the actual geometry is through the value of the five-dimensional Newton constant $G_5$. This is proportional to the volume of the internal manifold $H$ \cite{Witten:1998zw}. Thus $\langle T^{00}\rangle_{\beta}\sim {\rm Vol}(H)^{-1}$, in agreement with \eqref{ratio}.

The universal formula  \eqref{formula}, relating the energy density to the central charge, is implicit from the thermodynamics of black holes  combined with holographic considerations (see {\it e.g.} \cite{Kovtun:2008kw}). It is interesting to see that in our context it emerges from a mix of a number of nontrivial ingredients, including the OPE, a superconformal Ward identity and an  holographic computation of the thermal two-point function.

Remarkably, \eqref{georxt} can be resummed to give

\begin{equation}
\langle O(0) O(\tau,\vec{x})\rangle =   \frac{1}{|x|^{2\Delta}}\,e^{\frac{ \Delta\, \pi^4\,T^4}{120}\,C^{(1)}_2(\eta)\,|x|^4}\ ,
      \label{georxtresummed}
\end{equation}
with
$$
C^{(1)}_2(\eta) =-1 + 4 \eta^2 = \frac{3 \tau ^2-\vec x^2}{\tau ^2+\vec x^2}
$$
This completes the partial result \eqref{georresummed}, now including the full spacetime dependence. It exhibits an extremely interesting exponentiation of the energy-momentum tensor block. This exponentiation is similar to that observed for 4-point functions in CFT$_2$ \cite{Zamolodchikov:1985ie,Fitzpatrick:2014vua,Besken:2019jyw} and in CFT$_4$ \cite{Fitzpatrick:2019efk}.
However, in the present case of two-point functions, we have an explicit identification of the exponentiated block. Note that expanding the exponential we could read-off the coefficients for the corresponding blocks (each corresponding to suitably contracted powers of the energy-momentum tensor), thus providing an infinite sequence of predictions for such coefficients (see also \cite{Karlsson:2021duj,Kulaxizi:2019tkd} for related discussions).

\subsection{The $d=2$ case}

In order to gain further insight on the exponentiation, it is instructive to consider the case of 2d CFT's. The thermal two-point correlation function can also be computed in terms of  geodesics in two dimensions.
Let us begin by considering the simplest case of including only time dependence, {\it i.e.} we consider  $\langle O(0) O(\tau)\rangle $.
In euclidean signature, one finds that the geodesic (to be precise, each of its half-arcs) which goes from $-\tau_1$ to $\tau_1$ in the boundary is given by

\begin{equation} 
\tau=\frac{i}{2}\,\log\Big[ \frac{\cos \tau_1-\sqrt{z^2-\sin^2\tau_1}}{\cos \tau_1+\sqrt{z^2-\sin^2\tau_1}}\Big]\,.
\end{equation}
To compute the correlator we consider these two arcs hitting the boundary at $\tau_1$ and $-\tau_1$, respectively, and meeting at a bulk point over which we integrate by the saddle-point method. The  turning point is at the symmetric point $\tau=0$ and at $z=\sin \tau_1$. Then a small computation gives the following  2-point function:  

\begin{equation}
\langle O(0) O(\tau)\rangle =  \frac{\pi^{2\Delta}\,T^{2\Delta}}{\Big(\sinh (\pi T i\tau)\Big)^{2\Delta}}\,.
\end{equation}
One can similarly incorporate the $x$-dependence (see appendix A). In this way, one recovers the  general formula

\begin{equation}
\langle O(0) O(\tau,x)\rangle =  \frac{\pi^{2\Delta}\,T^{2\Delta}}{\Big(\sinh (\pi T (x-i\tau))\,\sinh (\pi T (x+i\tau))\Big)^{\Delta}}\,.
\label{finaldosd}
\end{equation}
which agrees with the well-known  expression for the thermal 2-point function in 2d CFT.
Note that in the present case of two dimensions the geodesic approach gives the exact correlation function for any $\Delta $.

Let us now study the limit leading to exponentiation in the present $d=2$ case.
Now the appropriate limit requires $T|x|\rightarrow 0$, $\Delta\to\infty $, with fixed
 $\Delta T^2\,|x|^2$. One finds

\begin{equation}
\langle O(0) O(\tau,x)\rangle= \frac{e^{\frac{\pi^2\,T^2}{3}\Delta\,(2\eta^2-1)\,|x|^2}}{|x|^{2\Delta}}\,.
\end{equation}
%

From these expressions, one can derive a formula for the thermal expectation value
of the energy-momentum tensor.
Consider the leading correction. We have $\frac{\pi^2\,T^2}{3}\Delta\,(2\eta^2-1)\,|x|^{2-2\Delta}$. We may re-write this as
\begin{equation}
    \lim_{d\rightarrow 2} \frac{1}{\beta^2}\,\frac{2\,\pi^2\,\Delta}{3}\,\frac{2!}{2^2\,(\frac{d}{2}-1)_2}C^{(\frac{d}{2}-1)}_2(\eta)\,|x|^{2-2\Delta}\,.
\end{equation}
Comparing with the general formula in \eqref{generalformcorrelator}, one finds

\begin{equation}
   \langle T^{00}\rangle_{\beta}=-\,\frac{\pi^2}{6\,\beta^2}\,(2\pi c_{\mathcal{T}})\,.
\end{equation}
For a free boson one has $c_{\mathcal{T}}=\frac{1}{2\pi^2}$, giving $\langle T^{00}\rangle_{\beta}=-\frac{\pi}{6}\,T^2=-\frac{2\,\zeta(2)}{{\rm vol}(S^1)}\, T^2$, just as expected \cite{Iliesiu:2018fao}.

\section{Teleological diving into the black brane} \label{n-point}

An important problem in holography is to describe  physics of the black hole interior in terms of boundary
CFT correlation functions. 
In a recent paper  \cite{Grinberg:2020fdj}, 
it was argued that one can measure the proper time
 to the black hole singularity by examining the asymptotic behavior of 
 thermal expectation values of
large charge operators. The expectation value is induced by a gravitational coupling
of the form $\phi W^2$, where $W^2=W_{\mu\nu\rho\sigma}W^{\mu\nu\rho\sigma}$ is the square of the Weyl tensor,
which represents the first possible coupling in a derivative expansion.
It was found that it has  a  factor of the form
\begin{equation}
 \label{properlen}
    \langle O\rangle \propto e^{-m \ell}\ ,\qquad \ell =\frac{R}{d}(\pm i\pi +\log 4) \ .
\end{equation}
The parameter $\ell $ represents the renormalized proper length from the boundary to the singularity,
computed as
\begin{equation}
    \ell =R \lim_{\epsilon\to 0}\left( \int_\epsilon^\infty \frac{dz}{z\sqrt{1-z^d}} +\log \epsilon\right)\ .
\end{equation}
Thus the  dependence of $\langle O\rangle $ on the conformal dimension $\Delta\sim m$ 
provides  a measurement of $\ell$.

It is natural to expect that higher point thermal correlation functions may   codify 
more information on the black hole geometry, perhaps in an intricate way.
In this section we shall consider the thermal two-point function in the CFT and compute the leading correction involving graviton emission for large dimension operators.

We will consider scalar operators of large dimensions carrying a $U(1)$ global charge.
In this case the relevant part of the action for the dual bulk scalar field is 
\begin{equation}
    I = \frac{1}{16\pi G_N} \int d^5x  \sqrt{g} \left[ g^{\mu\nu}\partial_\mu\bar\phi \partial_\nu\phi+m^2\bar\phi\phi +\alpha \bar\phi \phi W^2 \right]\ ,\qquad \Delta\approx mR\gg 1\ .
   \label{unWW}
\end{equation}
The coupling $\bar\phi\phi W^2$, where $W^2$ is the square of the Weyl tensor, appears at leading order in the derivative expansion. It leads to a new interaction vertex, which implies the existence of a Witten diagram representing  the following correction to the two-point function $\langle O(0)\bar O(x)\rangle$:

    \begin{equation}
 \label{intefi}
   I= \alpha \int_{\rm bulk} \, G(x_b^{(1)},x)\, G(x_b^{(2)},x)\, W^2(x)\ .
  \end{equation}
In terms of $r=z^2$, in the black brane background, $W^2= \frac{72}{R^4}\frac{r^4}{r_0^4}$.

We shall consider the case corresponding to Green's functions of operators inserted at coincident spatial points but at different times $t_1$ and $t_2$. Moreover, as we will be interested in correlators of large scaling dimension, we can trade  each of the Green's functions $G$ in \eqref{intefi} by the exponential of the geodesic length in \eqref{nuevax}. Finally, the integral in \eqref{intefi} can be done through the saddle-point approximation. Choosing, with no loss of generality, $t_2=-t_1$, it is clear by symmetry that the saddle point of the $t$ integration will  again lie at the symmetric point $t=0$ and $\vec x=0$. The general expression for $t$ in terms of $z$ and  $\mu$ is given by \eqref{tedepex}.
Therefore the integral to compute is

\begin{equation}
   I= c\int \frac{dr}{r^3} \ G(x_1-x'(z)) G(x_2-x'(z)) r^4\ ,\qquad c=  72R\frac{\alpha}{r_0^4}\ .
\label{iizz}
\end{equation}
 Due to the presence of the extra $W^2$ term, the two geodesic arcs will now meet at a different $r_{\star}$,  which can be obtained by studying the extrema of the integral \eqref{iizz} with respect to $r$. To find it we follow the same strategy as in previous sections. To begin with, the integrand in \eqref{iizz} can be written as $e^{-iS_T}$, where 

\begin{equation}
\label{esetotal}
S_T=S(z,\mu_1)+S(z,\mu_2)+ i\log r\ .
\end{equation}
It should be noticed that $S_T$ is a function which depends on the joining  point $r$ both explicitly and implicitly through $\mu_{1,2}$ --~the two rescaled momenta of each geodesic arc in \eqref{iizz}~-- since the variation is performed at fixed $t_1$.\footnote{
The overall coefficient obtained by the saddle-point approximation is inexact because the logarithmic term in \eqref{esetotal} is not multiplied by a large coefficient. However, following \cite{Grinberg:2020fdj}, we will not be interested in this coefficient but rather in the leading functional dependence, which is correctly reproduced  by the action.} Then, similarly to the cases studied previously, the saddle-point equation is

\begin{equation}
\label{saddx}
    0=\frac{dS_T}{dr}=  \frac{\partial S_T}{\partial r}+\frac{\partial S_T}{\partial \mu_1}\frac{\partial \mu_1}{\partial r}+\frac{\partial S_T}{\partial \mu_2}\frac{\partial \mu_2}{\partial r}\, .
\end{equation}
The last two terms give an identical contribution, since $\mu_1=-\mu_2\equiv \mu$.
In order to find $\frac{\partial \mu }{\partial r}$ we  differentiate \eqref{tedepex} at $t=0$ and constant $t_1$. Substituting the result into \eqref{saddx} we obtain the following equation:
\begin{equation}
     0= \frac{ i m \sqrt{1-r \left(\mu ^2+r\right)}}{r \left(r^2-1\right)} +\frac{i}{r}\,.
\end{equation}
This equation has three roots for $r$, with the following asymptotic behavior at large mass:
\begin{eqnarray}
    r_{\pm} & =& \pm i \Delta+\frac{\mu^2}{2 }+\mathcal{O}(\frac{1}{\Delta})\ ,  
\nonumber\\
r_3 &=&  \frac{1}{2} \left(\sqrt{\mu ^4+4}-\mu ^2\right)+\mathcal{O}(\frac{1}{\Delta^2})\ .
\end{eqnarray}
We note that the leading term in $r_3$ is the same as the saddle point \eqref{saddlez} that appeared in the calculation of the two-point function by the geodesic approximation in  section \ref{secciontresdos}.
As a result, to leading order, this gives rise to the same action:
\begin{equation}
    e^{-iS_T(r_3)} =  4^{-\Delta} \left(\mu ^4+4\right)^{\frac{\Delta}2}\ .
\label{aquis}
\end{equation}
On the other hand, $r_{\pm}$ give rise to complex geodesics with a complex action,
\begin{equation}
    e^{-iS_T(r_{\pm})} =
\left(-\frac{\mu ^2}{4}\mp \frac{i}{2}\right)^{\Delta}\ .
 \label{aqui}
\end{equation}
 In the case of the geodesics that join at $r_3$, the parameter $\mu$ in \eqref{aquis} is again 
determined by the condition \eqref{tmux} as a function of $t_1$. 

Now consider the contribution from the saddle points at  $r_{\pm}$. 
Substituting $r_{\pm}$ in \eqref{tedepex} and expanding for large $\Delta$, one finds

\begin{equation}
t_1=-i\tau_1 =\big(\mp \frac{1}{4}+\frac{i}{4}\big)\, \log \left(\frac{-\mu +(1\mp i)}{\mu
   +(1\mp i)}\right)+\mathcal{O}\left(\frac{1}{\Delta^2}\right)\ .
   \label{logui}
\end{equation}
This equation can be inverted to give
\begin{equation}
\label{mumux}
    \mu=-(1\pm i)\,\tan\big( (1\pm i)\,\tau_1\big)\,.
\end{equation}
This determines $\mu$ in terms of $\tau_1 $ for the saddle-point contribution at $r_\pm $, given by \eqref{aqui}.
Clearly, the contributions at $r_+$ and $r_-$ are complex conjugate and have equal weight $|e^{-i S_T}|$.

Let us  examine in detail the regime of validity of the approximation. Restoring $R$ and $z_0$,
\begin{equation}
    r_{\pm}=  \pm i z_0^2 \Delta  +\frac{\mu^2 z_0^4}{2R^2 }+\mathcal{O}\big(\frac{1}{\Delta}\big)\ .
\end{equation}
This shows that we must require $ \Delta \gg \frac{\mu^2 z_0^2}{R^2}$.
In turn,

\begin{equation}
\mu =  - (1\pm i)\frac{R}{z_0}\, \tan \left((1\pm i)\frac{ \tau_1}{z_0} \right)\ .
\label{mues}
\end{equation}
Then, the  condition $ \Delta \gg \frac{\mu^2 z_0^2}{R^2}$ requires,
\begin{equation}
     \big|\tan \left((1\pm i)\frac{ \tau_1}{z_0} \right)\big| \ll \sqrt{\Delta }\ .
\end{equation}
Since $\big|\tan \left((1\pm i)\frac{ \tau_1}{z_0} \right)\big|$ is bounded (in fact $<1.15$),
this condition is always satisfied for any time separation as long as $\Delta \gg 1$.

Having two contributions, $e^{-iS_T(r_3)}$ and $e^{-iS_T(r_{+})}+e^{-iS_T(r_-)}$, the dominant contribution
for a given $\tau_1$ is the one with largest modulus.
At small $\tau $, the dominant contribution is given by the saddle point sitting at $r_3$, since it will have large $\mu$ (see \eqref{taumu}) and therefore larger weight factor $|e^{-iS_T}|$. This gives the expected $1/|\tau|^{2m}$ behavior for nearly coincident points. 
However, 
there is a critical value of $\tau_1$ (namely $\tau_c\sim 1.16$) beyond which the dominant contribution is given by the saddle points at $r_{\pm}$, where the correction to the 2-point function is given by $e^{-iS_T(r_+)}+e^{-iS_T(r_-)}$.
\footnote{An open problem is understanding  the detailed physics, in particular the smoothness, of the transition between the small/large $\tau$ (see \cite{Dodelson:2020lal} for a recent discussion).}

Substituting \eqref{mumux} into \eqref{aqui}, we obtain
\begin{equation}
  e^{-iS_T(r_{\pm})} = {\rm const.}\,  2^{-\Delta}\, e^{\mp \frac{i\pi \Delta}{2}} \frac{1}{\left(\cos\left((1\pm i)\tau_1\right)\right)^{2\Delta}}\ .
\label{saddleuno}
\end{equation}
Note the factor $e^{-2 \Delta \ell}$, with $\ell=\pm \frac{i\pi \Delta}{2}+\frac12 \log 2$, 
representing renormalized length \eqref{properlen} to the black hole singularity.
This is the squared of the similar factor appearing in the one-point function in \cite{Grinberg:2020fdj}.
The reason of the squared is of course that this factor now accounts for two geodesics.
Thus we find that the information on the proper distance to the black hole singularity is
also present in a thermal two-point function. In addition, the $W^2$ correction to the correlation function has a non-trivial time dependence, which is
associated with the probability of graviton emission. 

Indeed, one can get a deeper insight on the meaning of this contribution to the correlation function by expanding
in exponential terms,

\begin{equation}
e^{-iS_T(r_{\pm})} \sim 2^{\Delta}\,e^{\mp i \frac{\Delta\pi}{2}}\,e^{-(1\mp i)\, \Delta\tau_1}\,\sum_{n=0}\frac{(-1)^n}{n!}\,\frac{(2\Delta+n-1)!}{(2\Delta-1)!}\,e^{-i\omega_n\, \tau_1}\, ,
\end{equation} 
with 

\begin{equation}
    \omega_n=(\Delta+2n)\,(\mp 1- i)\,.
\end{equation}

%

\noindent The complex frequencies $\omega_n $ coincide with the quasinormal frequencies of a scalar field in the black brane background  for $\Delta\gg 1$ \cite{Nunez:2003eq} (see also \cite{Fidkowski:2003nf,Amado:2008hw} for related discussions). 
The appearance of quasinormal frequencies seems to be associated with multi-graviton emission due to the coupling $\bar\phi \phi W^2$.

\section{Conclusions}\label{conclusions}

In this paper we have studied thermal 2-point functions for operators $O$ of large scaling dimension in four-dimensional CFT's with a holographic dual. The leading contribution to the 2-point function comes from a ``Witten diagram"\footnote{\textit{Stricto sensu} it is not a Witten diagram, as the propagators are not meeting at a vertex read off from the bulk theory lagrangian.} where one has two bulk-to-boundary propagators, one from a boundary point $x_b^{(1)}$ to a bulk point $x$ and another one from $x$ to a boundary point $x_b^{(2)}$, and the point $x$ where they meet is integrated  over the whole bulk. 
In the limit of large scaling dimension one can use the geodesic approximation and trade each propagator by a geodesic arc with the given endpoints and perform the integration over $x$ through a saddle-point approximation \cite{Dobashi:2002ar,Dobashi:2004nm,Janik:2010gc}. Even though this represents a big simplification, one is still left with the problem of writing the on-shell action for each geodesic arc as a function of the endpoints. The relevant equations are typically transcendental equations which cannot be inverted explicitly; consequently, the correlation function is
determined parametrically by a set of equations. Explicit expressions can be obtained by a systematic  expansion in powers of $T|x|$ (in particular, within the regime of validity of the OPE in field theory). In the case of correlation functions with only $\tau$-dependence, 
we have computed the correlator \eqref{geor} up to the order $(T\tau)^{12}$, and it is easy to extend this expansion to any desired order.
The full spacetime dependence of the thermal correlation function can be studied  by
a perturbative expansion in powers of $T$. Following this approach, we explicitly found the expected expansion  of the correlator in terms of Gegenbauer polynomials anticipated in \cite{Iliesiu:2018fao}. 
In the limit that we considered,  all contributions have been identified to correspond to appropriately contracted powers of the energy-momentum tensor. In this expansion the term corresponding to the energy-momentum tensor itself is particularly interesting, as the coefficient can be compared against expectations, and we found perfect agreement. This represents a highly non-trivial check of our results. 

An important consequence of the present results is that the leading terms in the   expansion at $\Delta\gg 1$,  $T|x|\ll 1$
can be resummed into an exponential of the block of the energy-momentum tensor. Upon expanding the exponential, this provides a prediction for an infinite sequence of coefficients which should hold in any CFT with a gravity dual to leading order in $\Delta$. Among these, it is particularly interesting the one for the energy-momentum tensor, which implies the  formula \eqref{formula} relating the energy density and the central charge for a 4d CFT with a gravity dual. In a similar limit, the 2-point function of a CFT$_2$ admits the same exponentiation of the energy momentum block. In fact, we can see a similar phenomenon even in the free theory, although in that case the exponentiated block is different. The exponentiation of the block of the energy-momentum tensor appears to be reminiscent of the exponentiation of Virasoro blocks in 4-point functions in CFT$_2$ as in \cite{Zamolodchikov:1985ie} (see also \cite{Fitzpatrick:2014vua,Besken:2019jyw}). The exponentiation can be traced to the existence of a semiclassical limit (with $\Delta^{-1}$ playing the role of $\hbar$). It should be noted that the exponent has no coupling dependence in the limit.
While from the holographic point of view the exponential form admits a simple explanation, from a purely field-theoretic point of view (for instance, by explicit diagrammatic computation of the 2-point function) it is a highly non-trivial prediction of strong coupling. This is very similar to the recent large charge studies in the $O(N)$ model \cite{Arias-Tamargo:2019xld,Badel:2019oxl,Watanabe:2019pdh,Arias-Tamargo:2019kfr}: for correlation functions of large charge ($\sim$ dimension) operators a semiclassical expansion in $\hbar\sim \Delta^{-1}$ emerges, so that there is an automatic exponentiation (see also \cite{Hellerman:2018xpi}).

One of the most fascinating aspects of thermal correlation functions in a CFT is that they
can encode information of the black hole interior through the holographic correspondence.
In section 4 we have considered a massive bulk scalar field $\phi$, dual to a CFT scalar field operator $O$ of large dimension, which couples to the Weyl tensor through an interaction $|\phi|^2W^2$.
This is the first possible interaction in a derivative expansion assuming $U(1)$ symmetry.
This could be, for instance, the case of chiral primaries in $\mathcal{N}=4$ SYM, which are charged under a $U(1)_R$ symmetry. This vertex gives rise to a correction to the 2-point thermal correlation function
which is contributed by complex geodesics entering deeply into the black hole interior.
The correction carries  a phase factor of the form $e^{-2\Delta \ell}$, where $\ell $ is the proper length
to the singularity, which is twice the similar factor appearing in the one-point function computed
in the model of \cite{Grinberg:2020fdj}. In addition, it has a time-dependence which carries information about
quasi-normal frequencies, which seem to be associated with multi-graviton emission.
Clearly, it would be extremely interesting to decipher detailed aspects of quantum black holes hidden in more general CFT thermal correlation functions.

\section*{Acknowledgements}

D.R-G is partially supported by the Spanish government grant MINECO-16-FPA2015-63667-P. He also acknowledges support from the Principado de Asturias through the grant FC-GRUPIN-IDI/2018/000174 J.G.R. acknowledges financial support from projects 2017-SGR-929, MINECO
grant PID2019-105614GB-C21, and  from the State Agency for Research of the Spanish Ministry of Science and Innovation through the “Unit of Excellence María de Maeztu 2020-2023” (CEX2019-000918-M).

\begin{appendix}

\section{Thermal two-point correlation function  in $d=2$}\label{CFT2}

In this appendix we compute the thermal two-point correlation function in two-dimensional CFT
with general spacetime dependence. The solution for the geodesic is obtained by integrating
\eqref{txeom}. We find
\begin{eqnarray}
&& t-t_1=  -\frac{1}{2} \log \bigg(\frac{\mu ^2-\nu ^2-1+z^2
   \left(\mu ^2+\nu ^2+1\right)+2 i \mu 
   \sqrt{(1-z^2)(1+\nu^2z^2)-z^2\mu^2 }}{\left(1-z^2\right) \left(-\nu ^2+(\mu +i)^2\right)}\bigg),
\nonumber\\
 &&   x-x_1 =\frac{1}{2} \log \bigg(\frac{\mu ^2+\nu ^2 \left(2
   z^2-1\right)+1+2 i \nu 
   \sqrt{(1-z^2)(1+\nu^2z^2)-z^2\mu^2 }}{\mu ^2-(\nu -i)^2}\bigg),
\nonumber\\
 &&   S =\frac{i}{2} \Delta \log \left(\frac{2+z^2 \left(-\mu
   ^2+\nu ^2-1\right)+2 \sqrt{(1-z^2)(1+\nu^2z^2)-z^2\mu^2 }}{z^2}\right)\ .
\end{eqnarray}
At $z=0$, the expressions for $x$ and $t$ satisfy  the boundary condition $x=x_1,\ t=t_1$.
The second geodesic arc has a similar solution but with the boundary condition at $z=0$, $x=-x_1$, $t=-t_1$,
and $\mu\to -\mu$, $\nu\to -\nu$.
By symmetry, they meet at some point $z$ where $x=0$, $t=0$. The action is the same for both geodesics.
The point where they meet is determined by demanding that it is an extremum of the action, {\it i.e.} it satisfies
\begin{equation}
    \frac{\partial S}{\partial z} +\frac{\partial S}{\partial \mu} \frac{\partial\mu}{\partial z} +\frac{\partial S}{\partial \nu} \frac{\partial\nu}{\partial z} =0\ .
\label{minimoss}
\end{equation}
The partial derivatives $\partial_z\mu$ and $\partial_z\nu$ are obtained by differentiating 
the above expressions for $t$ and $x$ at the joining point $t=x=0$ and fixed $t_1, \ x_1$. This leads to a system of two linear equations with two unknowns. Substituting the (lengthy) solution into \eqref{minimoss}, we get the condition
\begin{equation}
    \frac{\sqrt{1-z^2-\nu ^2 z^4+z^2 \left(\nu^2-\mu ^2\right)}}{z \left(1-z^2\right)}=0\ .
\end{equation}
As expected, the point $z$ that extremizes the action corresponds to the turning point of the geodesic obtained by the union of the two geodesic arcs, where $\dot t$ and $\dot x$ diverge. 
We have followed this longer derivation through extremization of $S$ because this derivation also applies in the presence of a vertex
operator in the Witten diagram where the turning point is no longer smooth.
The solution is
\begin{equation}
  z^2_*=  \frac{\sqrt{\left(\nu^2-\mu ^2-1\right)^2+4 \nu ^2}-\mu ^2+\nu ^2-1}{2 \nu ^2}\ .
\end{equation}
This gives the following formula   at the extreme point:
\begin{equation}
   e^{-2iS(z_*)}= \sqrt{\left(\mu ^2-\nu ^2+1\right)^2+4 \nu ^2}\ .
\end{equation}

Substituting the value of $z_*$ into $t$ and $x$, one can solve the resulting equations for $\mu $ and $\nu$ in terms of the initial values $(\pm t_1,\pm x_1)$ at $z=0$.
 We find
\begin{equation}
\mu= -\frac{ \sin (2 \tau_1)}{\cos (2 \tau_1)-\cosh (2 x_1)}\ ,\qquad \nu=\frac{ i \sinh (2x_1)}{\cos (2 \tau_1)-\cosh (2 x_1)}\ ,\qquad \tau_1=it_1\ .
\end{equation}
These expressions can be substituted into $z_*$ to express the turning point in terms
of $\tau_1$ and $x_1$,
\begin{equation}
  z^2_* = \frac{1}{2\cosh ^2(x_1)} (\cosh (2 x_1)-\cos (2 \tau_1))\ \  .
\end{equation}
One can check that $z_*\in (0,1)$, that is, the turning point lies outside the horizon.
Finally, substituting these formulas for $\mu,\ \nu$ into $e^{-2iS}$, one finds the correlation function
\begin{equation}
\langle O(\tau_1,x_1)O(-\tau_1,-x_1) \rangle =2^{\Delta }\, \left(\cosh (2 x_1)-\cos (2 \tau_1)\right)^{-\Delta }\ .
\end{equation}
Restoring the $z_0$ dependence by $x_1\to x_1/z_0$,  $t_1\to t_1/z_0$ and using that $z_0=1/(2\pi T)$, we obtain \eqref{finaldosd}.

\end{appendix}

\end{document}